\documentclass[12pt,english]{iopart}
\usepackage[T1]{fontenc}
\usepackage[latin9]{inputenc}
\usepackage[a4paper]{geometry}
\usepackage{verbatim}
\makeatletter
\usepackage{iopams}
\usepackage{setstack}
\usepackage[numbers, sort&compress]{natbib}

\makeatother
\usepackage{babel}

\begin{document}
\title[on radiation]{Note on radiation from an accelerated point charge and non-inertial
observers}
\author{{\Large B Broda}}
\address{{\large Department of Theoretical Physics, University of \L{}{\large \'o}d\'{z},
Pomorska 149/153, 90-236 \L{}{\large \'o}d\'{z}, Poland}}
\ead{{\large bobroda@uni.lodz.pl}}
\begin{abstract}
A simple gedankenexperiment is proposed showing observability of radiation
by an observer comoving with an accelerated charge.
\end{abstract}
\vskip1cm

Does the observer comoving (coaccelerating) with a nearby accelerated
charge detect (electromagnetic) radiation? This is an interesting
and non-trivial question in classical electrodynamics. For example
in one of recent papers \cite{key-1} the author presents a detailed
analysis of this problem and suggests a ``negative'' answer (no radiation).
The issue seems to be very controversial and completely different
answers can be found in literature (e.g.\ see references in \cite{key-1}).

In this note we present a simple argument showing that the comoving
observer, practically any observer (see the last paragraph of this
note), does detect radiation from an accelerated charge. Since our
analysis is given in the form of a gedankenexperiment, it should be
easily accessible even to novice students.

Our gedankenexperiment involves the 2 parties: Alice and Bob. Alice
is an inertial observer and rests in the inertial frame $F$. Bob
is a non-inertial observer and can travel by rocket. The engine of
the rocket can work entirely automatically according to a program
prescribed in advance. Before the experiment proper begins Alice and
Bob agree and prepare the program controlling the rocket engine. The
program should guarantee desired acceleration of Bob\textquoteright{}s
rocket. Therefore it should take into account not only inertia but
also forces exerted by the radiation from the accelerated charge $Q$.
(For any body of non-zero size there are forces due to the pressure
of radiation from $Q.$) No calculational details are necessary in
our reasoning but only the fact that the prepared program has to take
into account the radiation.

In the course of the experiment proper Alice can confirm that the
motion of Bob's rocket proceeds in accordance with the premises (provided
the program controlling the rocket engine is free of errors). Instead
Bob meets some difficulties. He must feel some additional forces (pressure)
acting on his rocket, besides thrust, because otherwise there would
be a contradiction between the preprogrammed work of the engine (taking
into account radiation present in the frame $F$) and the motion observed
by Bob. Such a contradiction would follow, for example, from the confrontation
of current measurement of Bob's velocity versus fuel burned with the
velocity calculated with the Tsiolkovsky rocket equation. Since there
is nothing more in space than radiation, the only possible interpretation
of this pressure is given in terms of the radiation.

The ``positive'' conclusion (presence of radiation) following from
the proposed gedankenexperiment is not confined to a particular relation
between the acceleration of the charge $Q$ and the acceleration of
the observer (Bob). Actually it follows from our considerations that
the accelerated observer (Bob) does always observe radiation provided
that he is within the range of radiation according to the inertial
observer (Alice).


\begin{thebibliography}{References}
\bibitem{key-1}Leonov A B 2012 Radiation from an accelerated point
charge and non-inertial observers \emph{Eur.\ J.\ Phys.}\ \textbf{33}
243--51 \end{thebibliography}
\end{document}